\documentclass[]{spie}  %>>> use for US letter paper
\usepackage{amsmath,amsfonts,amssymb}
\usepackage{graphicx}
\usepackage{wrapfig}
\usepackage[colorlinks=true, allcolors=blue]{hyperref}

\title{Liger at Keck Observatory: Design of Imager Optical Assembly and Spectrograph Re-Imaging Optics}

\author[a,b]{James Wiley}
\author[b]{Aaron Brown}
\author[c]{Renate Kupke}
\author[a,b]{Maren Cosens}
\author[a,b]{Shelley A. Wright}
\author[d]{Michael Fitzgerald}
\author[d]{Chris Johnson}
\author[e]{Tucker Jones}
\author[f]{Marc Kassis}
\author[d]{Evan Kress}
\author[d]{James E. Larkin}
\author[d]{Kenneth Magnone}
\author[f]{Rosalie McGurk}
\author[b]{Nils Rundquist}
\author[d]{Eric Wang}
\author[f]{Sherry Yeh}

\affil[a]{Department of Physics, University of California San Diego, USA}
\affil[b]{Center for Astrophysics and Space Sciences, University of California San Diego, USA}
\affil[c]{Department of Astronomy \& Astrophysics, University of California Santa Cruz, USA}
\affil[d]{Department of Physics \& Astronomy, University of California Los Angeles, USA}
\affil[e]{Department of Physics, University of California Davis, USA}
\affil[f]{W.M. Keck Observatory, Waimea, HI}

\begin{document} 
\maketitle

\begin{abstract}
Liger is an adaptive optics (AO) fed imager and integral field spectrograph (IFS) designed to take advantage of the Keck All-sky Precision Adaptive-optics (KAPA) upgrade for the W.M. Keck Observatory. We present the design and analysis of the imager optical assembly including the spectrograph Re-Imaging Optics (RIO) which transfers the beam path from the imager focal plane to the IFS slicer module and lenslet array.  Each imager component and the first two RIO mechanisms are assembled and individually aligned on the same optical plate. Baffling suppresses background radiation and scattered light, and a pupil viewing camera allows the imager detector to focus on an image of the telescope pupil. The optical plate mounts on an adapter frame for alignment of the overall system. The imager and RIO will be characterized in a cryogenic test chamber before installation in the final science cryostat.
\end{abstract}


\section{INTRODUCTION}
\label{sec:1}  % \label{} allows reference to this section

Liger is an adaptive optics (AO) fed imager and integral field spectograph (IFS) for the Keck I telescope at the W.M. Keck Observatory atop Maunakea in Hawaii. Liger will have larger fields of view, higher spectral resolving power ($R\sim 8,000-10,000$), and wider wavelength coverage ($0.8-2.4$ $\mu$m) than any current IFS \cite{Wright_2022}. Taking advantage of the improved observational capabilities provided by the Keck All-sky Precision Adaptive-optics (KAPA) upgrade to the current AO system, Liger will be a revolutionary instrument for a broad range of science cases\cite{Wizinowich_2020}\cite{Lu_2020}.

Liger makes use of a sequential imager and IFS design. The optical path for the spectrograph uses pick-off mirrors placed above the imager detector that allow the spectrograph and imager to be used concurrently in every mode\cite{Wright_2022}. The imager is the first major subsystem in the overall Liger instrument, placed directly after the cryostat entrance window.  It refocuses light onto the imager detector and pick-off mirrors as well as contains the pupil mask, filter wheel, and pupil viewing camera for the full system\cite{Cosens_2020}\cite{Cosens_2022}. Because the imager feeds the IFS subsystem, fabrication and alignment is crucial not just for the performance of the imager, but for the Liger instrument as a whole.

Imager optical components and spectrograph Re-Imaging Optics (RIO) are housed on the same optical plate. After all imager components are assembled and aligned, baffling is placed that blocks background radiation and scattered light from being seen by the imager detector and IFS subsystems. This baffling also provides a field stop at the AO focal plane which limits the field of view entering the optical system. A pupil viewing camera is placed 545.5mm away from the pupil mask which allows the imager detector to focus on the telescope pupil plane for alignment purposes.  The mechanism for this camera flips a single lens in-and-out of the beam path. The optical plate rests on an adapter frame that aligns the overall assembly within the imager test cryostat and integration into the science cryostat.

The RIO refocuses the beam from the imager focal plane to the IFS slicer and lenslet subsystems. The RIO consists of three mechanisms for rotating air-spaced doublets in-and-out of position and five fold mirrors along the beam path. Two of the three RIO mechanisms are mounted directly above the imager detector on the optical plate. The RIO allows selection of IFS plate scales for the slicer and lenslet subsystems, and in addition serves as to block light to each IFS assembly. The RIO allows selection between the 14 and 31 mas modes for the lenslet and the 75 and 150 mas modes for the slicer.

A vibrationally suppressed, cryogenically cooled, vacuum chamber was designed to characterize the Liger imager and RIO. This chamber was enlarged from a previous design \cite{Wiley_2020} to fit both the imager optical assembly and first two RIO mechanisms. The chamber consists of a steel body and aluminum lid and contains an aluminum cold shield separated from the body of the chamber with G10 A-frames. A custom made cart houses the vacuum, cryogenic, and electrical systems that support this experimental setup.

The remainder of this paper is organized as follows: \S\ref{sec:2} covers the Liger imager optical assembly and focuses on the design and analysis of the baffling and pupil viewing camera. The RIO is explained in more detail in \S\ref{sec:3}, and updates to the Liger imager test chamber are covered in \S\ref{sec:4}. A summary and future work section is given in \S\ref{sec:5}.

\section{Imager Optical Assembly}
\label{sec:2}

\begin{figure}[b]
   \begin{center}
   \includegraphics[width=0.8\textwidth,angle=0,trim={0in 0in 0in 0in},clip]{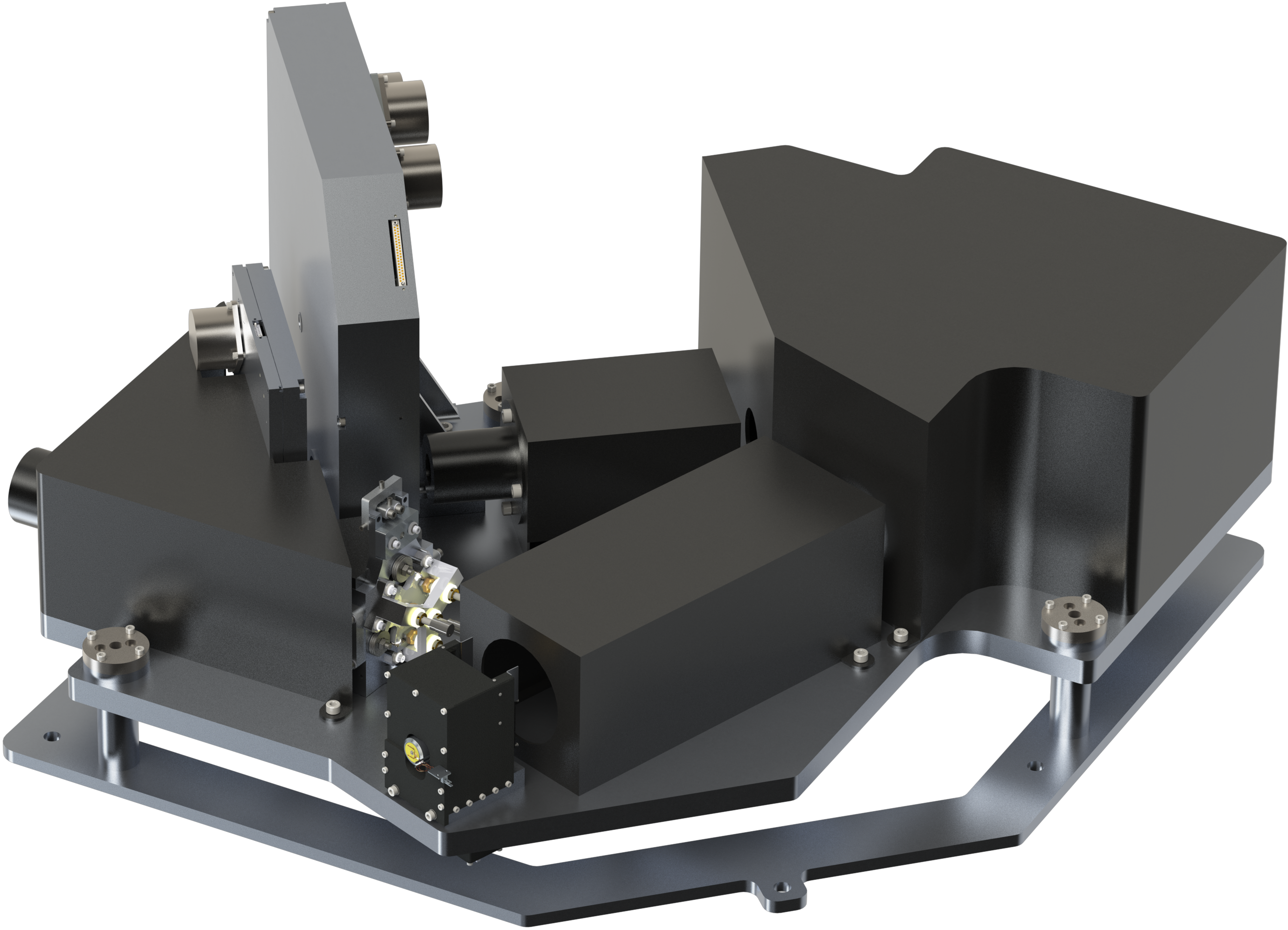}
   \end{center}
   \caption[Figure1]
  {
SolidWorks rendering of the Liger imager optical assembly without the RIO. The imager optical plate with the pupil wheel, filter wheel, detector, and baffling are shown. The first off-axis parabola (OAP) is partially covered and the second OAP, flat mirror, and pupil viewing camera are fully covered by baffle boxes. The optical plate rests on an adapter frame via three canoe spheres that is used to align the imager in the testing cryostat and science cryostat.}
\label{fig:1}
\end{figure}

The Liger imager is custom designed but takes heritage from Keck OSIRIS \cite{Larkin_2006}. It is optimized for low wavefront error and high throughput\cite{Wright_2020}. It provides a 20.5"x20.5" field of view with 10 mas spatial sampling. Thanks to Liger's sequential design, the imager is used in parallel with the IFS in all observing modes. It is the first major optical assembly along the beam path in the Liger instrument and is located directly behind the entrance window. The simple optical design uses two Off-Axis Parabolic (OAP) mirrors to transfer the beam path from the AO focal plane to the imager detector.

The imager optical assembly consists of two off-axis parabolas (OAPs), a flat mirror, a pupil wheel, a filter wheel, a pupil viewing camera, and Teledyne Hawaii-2RG detector. These optical assemblies are mounted to the same light-weighted optical plate as the first two RIO mechanisms. Each component can be aligned on the optical plate individually. An adapter frame provides adjustment for the optical platform as a whole. For a more detailed overview of the pupil wheel and filter wheel see Cosens et al. 2020 \cite{Cosens_2020}, and for a more detailed overview of the detector mount see Cosens et al. 2022 \cite{Cosens_2022}. A rendering of the full imager optical assembly is given in Figure \ref{fig:1} showing the first OAP mount, the pupil wheel, filter wheel, detector mount, baffling, and adapter frame.

The adapter frame rests on the cold shield base and has three pedestals for mounting to the optical plate. The optical plate uses shims for height adjustment and lowers on to canoe spheres that rest on top of the adapter pedestals that will allow for positional repeatability. The adapter frame has two tabs that allow for $\pm 4$mm of movement which is within the tolerance stack-up of the imager test cryostat. Fine threaded M5 set screws are used for this adjustment which gives a precision of a quarter screw turn of 125$\mu$m.

After installation of each major optical component, baffling is installed to cover all optics and the imager beam path. The baffling consists of eight major components: four baffle boxes, two baffle tubes, and two shrouds that cover the OAPs. Baffling bolts to the imager optical plate and sufficiently suppresses scattered light. \S\ref{sec:2.1} shows the baffling design in more detail and presents the scattered light analysis performance.

The pupil viewing camera mechanism moves in-and-out of the beam path, which allows the imager detector to focus on the telescope pupil for alignment.  This is achieved with a single lens placed 545.5mm away from the pupil mask. This lens is mounted on a flip mechanism similar to the RIO. \S\ref{sec:2.2} shows the preliminary design and analysis in Zemax for the pupil viewing camera.

\subsection{Baffling}
\label{sec:2.1}

\begin{figure}[b]
\begin{center}
\minipage{0.489\textwidth}
\begin{center}
  \includegraphics[width=0.99\linewidth,trim={0.0in 0.0in 0.0in 0.0in}]{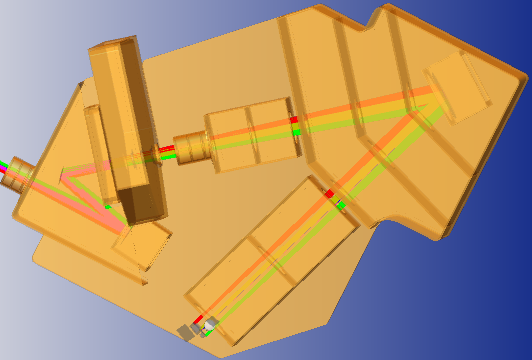}
  \end{center}
\endminipage\hfill
\minipage{0.509\textwidth}
\begin{center}
  \includegraphics[width=.99\linewidth,trim={0.0in 0.0in 0.0in 0.0in}]{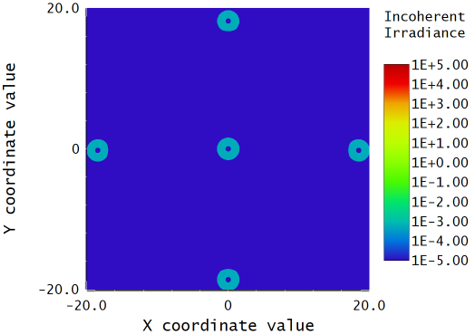}
\end{center}
\endminipage\hfill
\vspace{0.25cm}
\caption[Figure2]{Left: Zemax model of the Liger imager baffling including the optical plate, pupil wheel, and filter wheel. Right: Ray trace analysis with $5\times10^5$ rays for each of five imager field points. The detector shows each field point at the $3\times10^4$ Watts/cm$^2$ level with ghost donuts at $<10^{-3}$ Watts/cm$^2$. These ghosts are due to internal reflection off the cryostat entrance window, and all other ghosts are successfully suppressed by baffling.}
\label{fig:2}
\end{center}
\end{figure}

The baffling is a critical component of the overall imager design as it blocks scattered light and background radiation from the imager detector and sequential IFS. The current design uses 8 gauge 6061 T6 aluminum sheet metal that is folded and welded together. There is a 0.5" wide lip along the bottom of the baffling that will provide stability and be used to fasten the baffling assembly to the optical plate. Bolting to the optical plate provides sufficient positional tolerance.

The baffling covers all optical components on the imager with only small gaps after the filter wheel, between the baffling boxes, and before the detector. Baffling holes are cut in the sheet metal for the beam path which are small enough to block outside light while large enough to avoid vignetting on the detector. Multiple faces inside the baffling suppress specular reflection. The baffling will be painted a special black that is sufficient for cryogenic operation to further suppress reflections.

Only two of the baffle boxes contain components under them. This allows for easier access to the imager components by removing a smaller baffle box rather than larger baffling as a whole.  This also minimizes interferences with other components on the imager optical assembly and in the science cryostat as there are already necessary cutouts on parts of the baffling.

The first baffle tube bolts to the first baffle box and contains the field stop for the assembly. The second baffle tube bolts to the second baffle box and allows the baffling to be placed closer to the filter wheel exit. The two baffle boxes contain a bolt hole pattern and PEM nuts are pressed into these holes. The first OAP and flat mirror are covered by the first baffle box and the second OAP and the pupil viewing camera are covered by the large, third baffle box.

A simplified version of the baffling including the imager optical plate and the pupil and filter wheel was included in a Zemax model of the imager to analyze the scattered light seen by the detector. Figure \ref{fig:2} shows this simplified model and the ray trace analysis. The baffling allows the full field of view to reach the detector without vignetting. The resultant ghosting is due to internal reflections off the cryostat entrance window and can only be mitigated with a high quality window and Anti-Reflection (AR) coating.  The detector is tilted $1^{\circ}$ and the filters are tilted $3^{\circ}$ to suppress reflections, the remainder are suppressed by the baffling.

\subsection{Pupil Viewing Camera}
\label{sec:2.2}

\begin{figure}[b]
\begin{center}
\minipage{0.489\textwidth}
\begin{center}
  \includegraphics[width=0.99\linewidth,trim={0.0in 0.0in 0.0in 0.0in}]{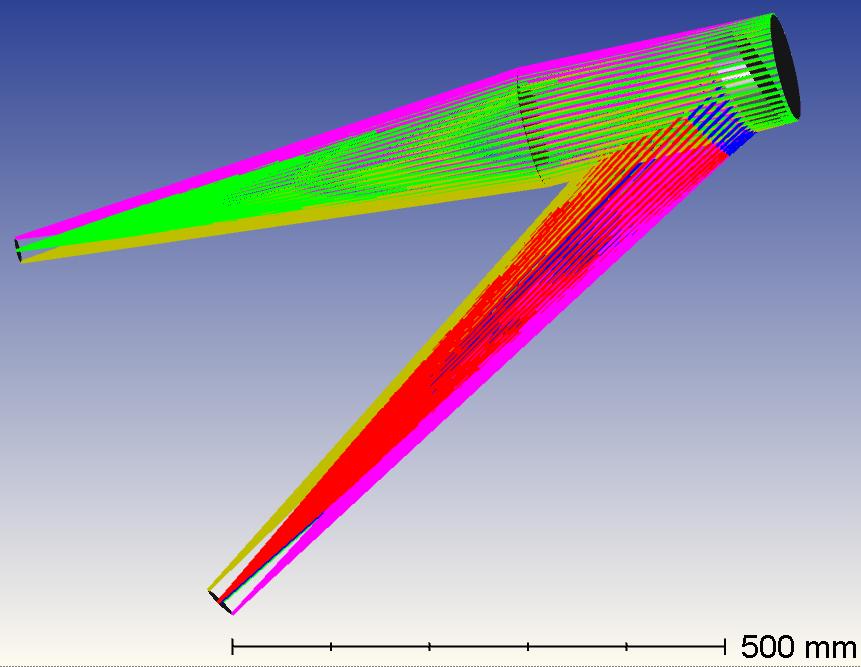}
  \end{center}
\endminipage\hfill
\minipage{0.509\textwidth}
\begin{center}
  \includegraphics[width=.99\linewidth,trim={0.0in 0.0in 0.0in 0.0in}]{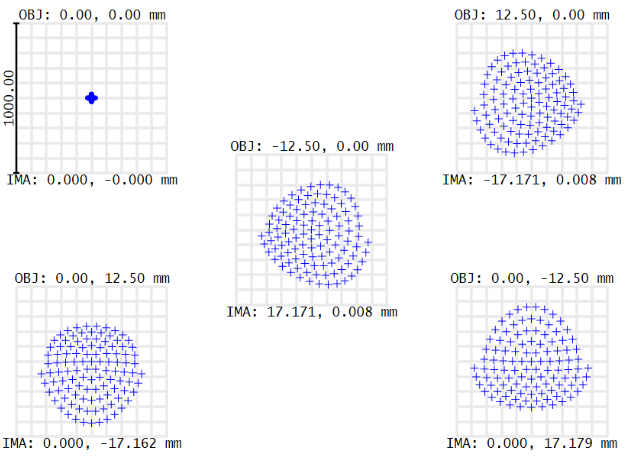}
\end{center}
\endminipage\hfill
\vspace{0.25cm}
\caption[Figure3]{Left: Zemax shaded model of the pupil viewing camera from the pupil mask to the imager detector. The CaF$_2$ lens is placed 545.5mm from the pupil image plane. Right: Spot diagram from Zemax showing the performance of the pupil viewing camera at each of five field points. Each field shown is 1000 $\mu$m across. The RMS radius of the spot size is about 12$\mu$m ($\sim0.5$ pixels) at the center and 250$\mu$m ($\sim12$ pixels) near the detector edges.}
\label{fig:3}
\end{center}
\end{figure}

The pupil viewing camera is a necessary component of the Liger imager optical assembly. It is located between the pupil wheel and second OAP and focuses an image of the telescope pupil onto the imager focal plane. The pupil viewing camera accomplishes this by simply moving a single calcium fluoride (CaF$_2$) lens in and out of the beam path. The magnification of this lens and the OAP combined must be less than 1.412 to avoid overfilling the imager detector.

\begin{wrapfigure}{r}{0.45\textwidth}
\vspace{0.15in}
  \begin{center}
    \includegraphics[width=0.43\textwidth]{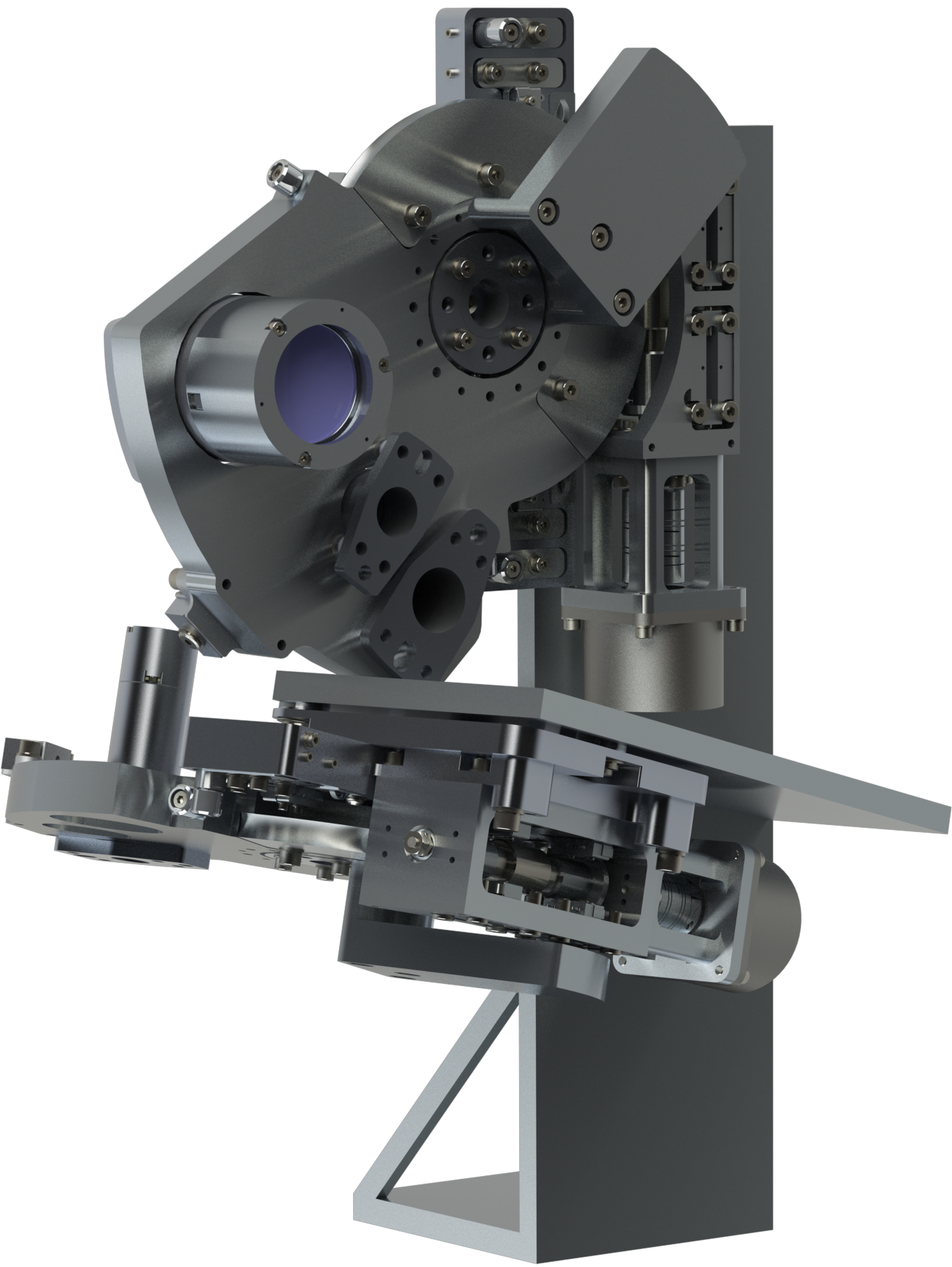}
  \end{center}
  \caption[Figure4]{SolidWorks rendering of the first two RIO mechanisms attached to the bracket for alignment on the optical plate. This view does not show the fold mirrors between the two mechanisms.}
\label{fig:4}
\end{wrapfigure}

An optical model for the pupil viewing camera was created in Zemax and an optimization routine was run to find the ideal placement for the lens at a wavelength of 2$\mu$m. For a CaF$_2$ lens at a temperature of 77 K, it was found the ideal placement is 545.5mm from the pupil plane. The shaded model of this is shown in Figure \ref{fig:3} as well as a spot diagram detailing the optical performance of the pupil viewing camera at each of five field points on the imager detector. The RMS spot size of the pupil viewing camera is about half a pixel or 12$\mu$m in radius at the center of the field and 12 pixels or 250$\mu$m in radius at the edges of the imager detector. Further analysis will be completed to determine whether this is sufficient or if this spot size can be reduced.

While the pupil viewing camera has not been fully designed, it will use the same flip mechanism as the RIO to switch the lens in-and-out of position. The lens does not need to be as precisely placed as the other optical components, so a simple mount design will suffice. The pupil viewing camera is located in the large baffle box on the imager optical plate and the design may change slightly to accommodate the camera. The pupil viewing camera will be used during alignment of the optical components of the imager, and when the imaging camera is integrated with the IFS in the science cryostat.

\section{Re-Imaging Optics}
\label{sec:3}

The spectrograph RIO houses three mechanisms that use simple re-imaging optics, two of which are placed directly above the imaging detector, to preserve the pristine image quality provided by KAPA\cite{Wright_2020}. It transfers the beam from the imager focal plane to either the lenslet or slicer subsystems. As well as choosing between the lenslet and slicer, the RIO also allows choosing between the 14 and 31 mas mode for the lenslet and the 75 and 150 mas mode for the slicer.

In total, the RIO contains six air-spaced doublets which are mounted on the three separate mechanisms as well as up to five fold mirrors along the beam path. For the lenslet modes, the beam passes through a single doublet, and both slicer modes pass through two doublets. The first mechanism is placed vertically above the detector and contains the doublet for the 14 mas lenslet mode. The first RIO mechanism also blocks the beam from entering the not chosen spectrograph path. After it, fold mirrors reflect light horizontally into the second mechanism which contains doublets for the 31 mas lenslet mode and the 75 and 150 mas slicer modes. 

The third mechanism is further down the beam path and contains a second pair of doublets for the 75 and 150 mas slicer modes only. Fold mirrors send light from the third mechanism to the image slicer. For the lenslet path, fold mirrors feed the light to the lenslet array after the second RIO mechanism.

The first two of the three RIO mechanisms are aligned relative to each other on a bracket that is then aligned on the imager optical plate. These two mechanisms will be tested along with the imager in the test cryostat. Figure \ref{fig:4} shows a SolidWorks rendering of the first two RIO mechanisms mounted on this bracket. The view shown does not include the two fold mirrors between the first and second RIO mechanisms.

\section{Experimental Setup}
\label{sec:4}

\begin{figure}[t]
   \begin{center}
   \includegraphics[width=1\textwidth,angle=0,trim={0in 0in 0in 0in},clip]{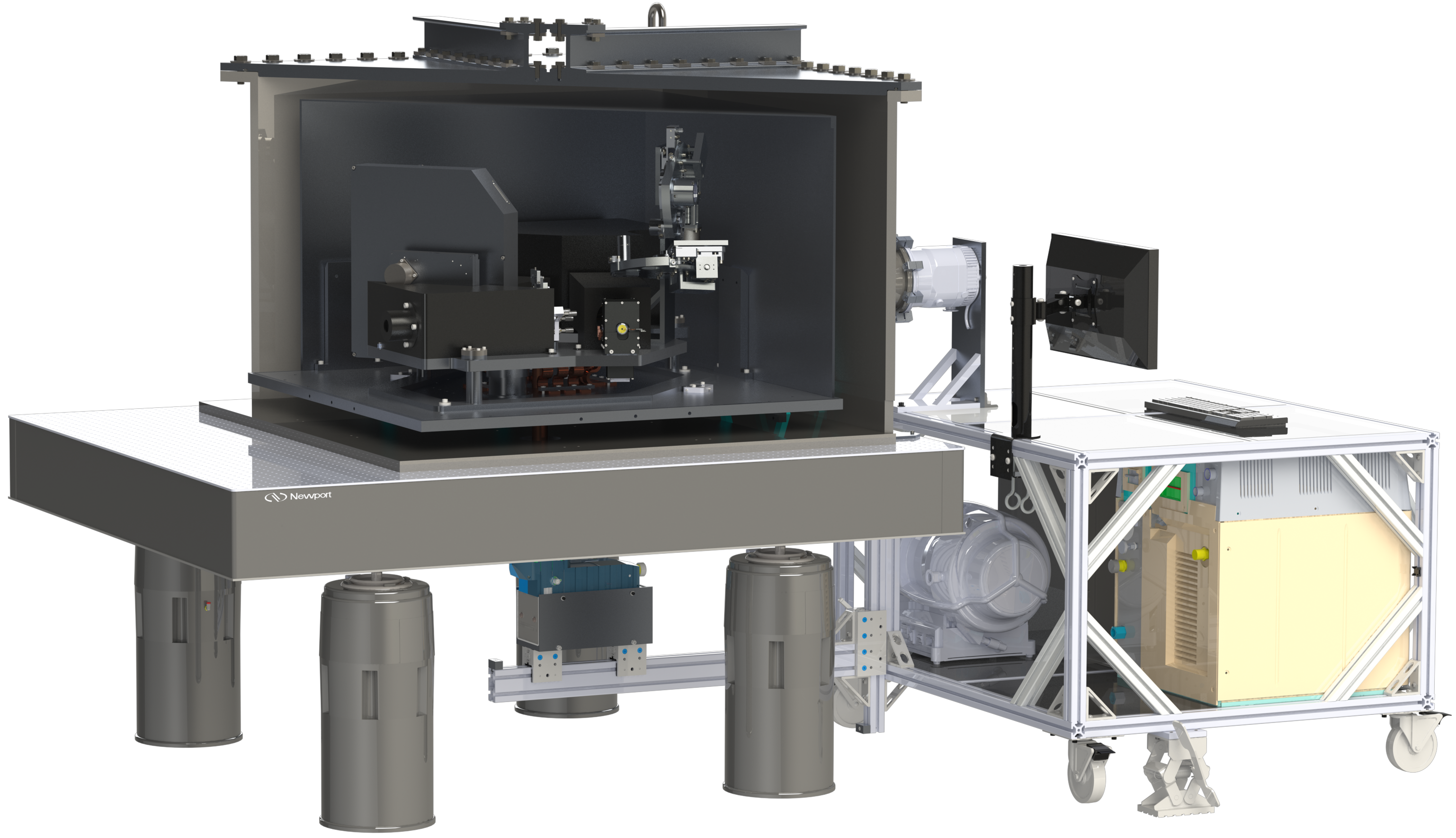}
   \end{center}
   \caption[Figure4]{SolidWorks rendering of the Liger imager test cryostat experimental setup. This view shows cuts in the vacuum chamber and cold shield walls to show the imager installed within the test cryostat. The setup is vibrationally isolated using a Newport isolation system and is serviced by a custom made cart that houses the vacuum, cryogenic, and electrical components of the setup.}
\label{fig:5}
\end{figure}

The experimental setup, as shown in Figure \ref{fig:5}, consists of an AISI 304 stainless steel vacuum chamber that houses a 6061 T6 cold shield. The optical assembly rests and is adjusted on the base of the cold shield. The cold shield is thermally isolated from the chamber base with G10 A-frames. Multi-Layer Insulation (MLI) is used between the cold shield and chamber walls to reduce heat transfer through radiation. The top shell of the cold shield lowers over the optical assembly and bolts to guides along the edges. The vacuum chamber rests on a Newport vibration isolation system with a passive damped optical table on active pneumatic isolators. A custom made cart houses the cryogenic, vacuum, and electrical components of the setup. The experimental setup operates below a pressure of $10^{-5}$ Torr and temperature of 77 K with a maximum deflection between two points due to vibration of 1 $\mu$m. 

The interior working dimensions of the setup is 1053$\times$1053$\times$670mm which is large enough to fit the full imager optical assembly as well as the first two RIO mechanisms. This gives a minimum half an inch of clearance around the full assembly when installed. The cold head raises up into the chamber, and a copper extension rises through a hole in the middle of the cold shield. Cold straps clamp to the extension and the cold shield base. Getters are placed near the cold straps to absorb condensation during cool down. 

A previous version of the vacuum chamber was designed and analyzed in Wiley et al. 2020 \cite{Wiley_2020}. This chamber was enlarged to fit the first two RIO mechanisms in addition to the imager optical assembly. The larger chamber has a 1” thick base and lid to withstand the larger force on it from the vacuum. The walls of the chamber are 5/8" thick and welded together and onto the base. A top flange, 3” wide and 1" thick, is welded on top of the chamber walls to allow the lid to lower down and form a vacuum seal against its O-ring surface. From a SolidWorks simulation, the maximum stress in the chamber is 104 MPa which is about a factor of two below the yield strength of AISI 304, and the maximum deflection is 1.4mm.

The vacuum chamber is positioned on the Newport optical bench with guides and bolted in place with brackets after installation. Earthquake straps may be used to further secure the chamber to the optical platform if needed. The vacuum and cryogenic lines are connected over bellows and vibration isolation pads to the chamber to reduce induced vibrations. Electrical feedthroughs, vacuum gauges, and other components are installed on NW-100 and NW-160 flanges along the chamber sides. The vacuum line is connected over a NW-160 flange and a gate valve. The CaF$_2$ entrance window is mounted on a NW-100 flange. The imager will be aligned inside the test cryostat to a telescope simulator which will be mounted on the optical bench in front of the entrance window.

The experimental setup will be located on a fourth floor laboratory on the University of California San Diego campus.  It will require a crane that can lift up to one ton and a clean room for assembling the components that are installed in vacuum. After characterization of the imager, it will be moved to the University of California Los Angeles where it will be installed in the final science cryostat. Analysis has been performed to ensure the assembly survives shipping between these two locations and the final location on Maunakea.

\section{Summary and Future Work}
\label{sec:5}

This paper describes the design of the imager optical assembly focusing on the baffling and pupil viewing camera as well as the spectrograph RIO system and the experimental test setup. The analysis included shows the baffling successfully suppresses internal reflections and background radiation. The optimal position for the pupil viewing camera is 545.5mm from the pupil plane for a CaF$_2$ lens at 77 K and provides a sufficient magnification and spot size at that location. The assembly as a whole will be tested in a vibration suppressed, cryogenic, vacuum chamber before being assembled in the final science cryostat.

Each major component of the imager optical assembly: the two OAPs, flat mirror, pupil wheel, filter wheel, detector, and pupil viewing camera, rest on a light-weighted optical plate. The first two RIO mechanisms also mount to this plate.  After alignment of the individual components, baffling is then installed. This optical assembly is lowered onto an adapter frame that allows the full optical system to then be aligned.

Future work for the baffling includes incorporating the Re-Imaging Optics into the overall design as the RIO rests directly above the detector and needs baffling of its own. Future work also includes finishing the design for the pupil viewing camera mechanism which will move the CaF$_2$ lens in and out of the beam path to allow the detector to focus on an image of the telescope pupil. The current design of this mechanism fits within the large baffling box.

The Liger imager optical assembly and RIO meet the requirements for the overall Liger system. The Liger imager and RIO sequentially feeds the Liger IFS and serves as its own unique science case for AO imaging.

\acknowledgments  
The Liger instrumentation program is supported by the Heising-Simons Foundation, the Gordon and Betty Moore Foundation, University of California Observatories, and W. M. Keck Observatory.

\bibliography{report}

\begin{thebibliography}{1}

\bibitem{Wright_2022}
{Wright}, S.~A. et~al., ``{Liger at Keck Observatory: Overall Design
  Specifications and Science Drivers},'' in [{\em Ground-based and Airborne
  Instrumentation for Astronomy IX}{\nolinebreak\hspace{0.1em}]},  SPIE (July
  2022).

\bibitem{Wizinowich_2020}
{Wizinowich}, P., {Chin}, J., {Correia}, C., {Lu}, J., {Brown}, T., {Casey},
  K., {Cetre}, S., {Delorme}, J.~R., {Gers}, L., {Hunter}, L., {Lilley}, S.,
  {Ragland}, S., {Surendran}, A., {Wetherell}, E., {Ghez}, A., {Do}, T.,
  {Jones}, T., {Liu}, M., {Mawet}, D., {Max}, C., {Morris}, M., {Treu}, T., and
  {Wright}, S., ``{Keck all sky precision adaptive optics},'' in [{\em Society
  of Photo-Optical Instrumentation Engineers (SPIE) Conference
  Series}{\nolinebreak\hspace{0.1em}]},  {\em Society of Photo-Optical
  Instrumentation Engineers (SPIE) Conference Series} {\bf 11448},  114480E
  (Dec. 2020).

\bibitem{Lu_2020}
{Lu}, J.~R., {Wizinowich}, P., {Correia}, C., {Chin}, J., {Cetre}, S.,
  {Lilley}, S., {Ragland}, S., {Wetherell}, E., {Birrer}, S., {Do}, T.,
  {Drechsler}, W., {Ghez}, A., {Jones}, T., {Liu}, M., {Mawet}, D., {Max}, C.,
  {Morris}, M., {Treu}, T., and {Wright}, S., ``{KAPA: A new Keck laser-guide
  star AO system that increases image quality and sky coverage},'' in [{\em
  American Astronomical Society Meeting Abstracts
  \#235}{\nolinebreak\hspace{0.1em}]},  {\em American Astronomical Society
  Meeting Abstracts} {\bf 235},  118.03 (Jan. 2020).

\bibitem{Cosens_2020}
{Cosens}, M., {Wright}, S.~A., {Arriaga}, P., {Brown}, A., {Fitzgerald}, M.,
  {Jones}, T., {Kassis}, M., {Kress}, E., {Kupke}, R., {Larkin}, J.~E., {Lyke},
  J., {Wang}, E., {Wiley}, J., and {Yey}, S., ``{Liger for Next-Generation Keck
  Adaptive Optics: Filter Wheel and Pupil Design},'' in [{\em Ground-based and
  Airborne Instrumentation for Astronomy VI}{\nolinebreak\hspace{0.1em}]},
  {Shields}, J., ed., {\em Society of Photo-Optical Instrumentation Engineers
  (SPIE) Conference Series} {\bf 11447},  11447--280 (Dec. 2020).

\bibitem{Cosens_2022}
{Cosens}, M., {Wright}, S.~A., {Brown}, A., {Fitzgerald}, M., {Johnson}, C.,
  {Jones}, T., {Kassis}, M., {Kress}, E., {Kupke}, R., {Larkin}, J.~E.,
  {Magnone}, K., {McGurk}, R., {Rundquist}, N.-E., {Sohn}, J.~M., {Wang}, E.,
  {Wiley}, J., and {Yeh}, S., ``{Liger at Keck Observatory: Imager Detector and
  IFS Pick-off Mirror Assembly},'' in [{\em Ground-based and Airborne
  Instrumentation for Astronomy IX}{\nolinebreak\hspace{0.1em}]},  {\em Society
  of Photo-Optical Instrumentation Engineers (SPIE) Conference Series}, SPIE
  (July 2022).

\bibitem{Wiley_2020}
Wiley, J.~H., Mathur, K., Brown, A.~M., Wright, S.~A., Cosens, M., Maire, J.,
  Fitzgerald, M.~P., Jones, T.~A., Kassis, M., Kress, E., and et~al., ``Liger
  for next-generation keck adaptive optics: design of opto-mechanical test
  chamber,'' {\em Ground-based and Airborne Instrumentation for Astronomy
  VIII}~{\bf 11447} (Dec 2020).

\bibitem{Larkin_2006}
{Larkin}, J., {Barczys}, M., {Krabbe}, A., {Adkins}, S., {Aliado}, T., {Amico},
  P., {Brims}, G., {Campbell}, R., {Canfield}, J., {Gasaway}, T., {Honey}, A.,
  {Iserlohe}, C., {Johnson}, C., {Kress}, E., {LaFreniere}, D., {Lyke}, J.,
  {Magnone}, K., {Magnone}, N., {McElwain}, M., {Moon}, J., {Quirrenbach}, A.,
  {Skulason}, G., {Song}, I., {Spencer}, M., {Weiss}, J., and {Wright}, S.,
  ``{OSIRIS: a diffraction limited integral field spectrograph for Keck},'' in
  [{\em Society of Photo-Optical Instrumentation Engineers (SPIE) Conference
  Series}{\nolinebreak\hspace{0.1em}]},  {McLean}, I.~S. and {Iye}, M., eds.,
  {\em Society of Photo-Optical Instrumentation Engineers (SPIE) Conference
  Series} {\bf 6269},  62691A (June 2006).

\bibitem{Wright_2020}
{Wright}, S.~A., {Larkin}, J.~E., {Jones}, T., {Aliado}, T., {Armus}, L.,
  {Brown}, A., {Chisholm}, E., {Cosens}, M., {Dekaney}, R., {Do}, T.,
  {Fassanacht}, C., {Fisher}, D., {Fitzgerald}, M., {Ghez}, A., {Hirtenstein},
  J., {Johnson}, C., {Kassis}, M., {Keane}, J., {Kelley}, P., {Kirby}, E.,
  {Konopacky}, Q., {Kress}, E., {Kupke}, R., {Lu}, J., {Lyke}, J., {Marley},
  M., {Medling}, A., {Millar-Blanchaer}, M., {Nash}, R., {Nierenberg}, A.,
  {Reddy}, N., {Rich}, M., {Ruffio}, J.-B., {Rundquist}, N.-E., {Sand}, D.,
  {Sanders}, R., {Sandstrom}, K., {Shapley}, A., {Sohn}, J.-M., {Surya}, A.,
  {Treu}, T., {Wang}, E., {Weber}, B., {Wiley}, J., {Wizinowich}, P., {Wong},
  M., {Yeh}, S., and {Zonca}, A., ``{Liger for Next-Generation Keck Adaptive
  Optics: Overall Design and Status},'' in [{\em Ground-based and Airborne
  Instrumentation for Astronomy VI}{\nolinebreak\hspace{0.1em}]},  {Shields},
  J., ed., {\em Society of Photo-Optical Instrumentation Engineers (SPIE)
  Conference Series} {\bf 11447},  11447--331 (Dec. 2020).

\end{thebibliography}
\bibliographystyle{spiebib}

\end{document}